\documentclass[3p,times,procedia]{elsarticle}
\usepackage{nupha_ecrc}

\volume{00}

\firstpage{1}

\journalname{Nuclear Physics A}

\runauth{P.B. Gossiaux et al.}

\jid{nupha}

\jnltitlelogo{Nuclear Physics A}

\usepackage{amssymb}
\usepackage[figuresright]{rotating}
\usepackage{here}
\begin{document}

\begin{frontmatter}



%% Instructions from Editor: Please use the following \dochead only in the preprint version (e-print arXiv etc.); 
%% use empty \dochead{} when submitting to Nuclear Physics A!
% \dochead{XXVIth International Conference on Ultrarelativistic Nucleus-Nucleus Collisions\\ (Quark Matter 2017)}
%% e.g. \dochead{17th International Conference on Dynamical Processes in Excited States of Solids}

\title{Global view on coupled dynamics of heavy and light flavor observables from EPOSHQ}


\author{P.B. Gossiaux$^1$, J. Aichelin$^1$, M. Nahrgang$^1$, V. Ozvenchuk$^{1,2}$ and K. Werner$^1$}
\address{$^1$SUBATECH, UMR 6457, IMT Atlantique, Universit\'e de Nantes, IN2P3/CNRS.\\ 4 rue Alfred Kastler, 44307 Nantes cedex 3, France}
\address{$^2$ Cracow, INP, ul. Radzikowskiego 152, 31-342 Krak\'ow, Poland}

\begin{abstract}
 We present an analysis of the comparison between 2nd and 3rd flow
 harmonics of $\pi$, protons and D mesons produced in ultrarelativistic 
 heavy ion collisions. We advocate that such an analysis could turn out
 to be a good strategy in order to quantify the off-equilibrium dynamics of heavy flavor at lower transverse momentum $p_T$.
\end{abstract}


\end{frontmatter}

% \linenumbers

\section{Introduction}
While heavy flavor (HF) mesons are generally accepted as being hard probes of the QGP formed in ultrarelativistic heavy ion collisions, 
some signs of -- at least partial -- thermalization for their parent heavy 
quarks (HQ) in the medium is often advocated for HF mesons formed at intermediate $p_T$ ($p_T\lesssim m_Q$)~\cite{Cao:2011}. In particular, the finite (so called) elliptic flow of HF mesons observed in semi-central collisions is often cited as a proof that the parent HQ benefit from the bulk flow, and hence thermalize in the
QGP. In~\cite{Nahrgang:2015}, we addressed the question of higher flow harmonics for both D and B mesons and for various centralities. We came up with the novel observation that heavier and heavier flavor mesons "benefit" less and less from the bulk flow, especially for larger and larger centralities and higher flow harmonics. This effect, partially confirmed by the $v_3(D)$ measurement by the the CMS collaboration~\cite{CMS:2016}, can be attributed to the inertia of HQ
which delays their response to the bulk flow.
At QM 2014, we have demonstrated that a cleaner analysis for this promising observables could be achieved by considering HQ flow harmonics as a function of the initial spatial eccentricities $\epsilon_n$ which are now understood (see refs 39-41 of~\cite{Nahrgang:2015}) as genuine seeds of the corresponding bulk flow. This is particularly relevant when event by event (EBE) fluctuations are included in state of the art descriptions of the bulk evolution like f.i. EPOS~\cite{EPOS:2010}. Although quite relevant for the physical understanding, $\epsilon_n$-dependent analysis cannot be easily compared with experimental data, as the $\epsilon_n$ cannot be directly measured.
The goal of this study is hence to investigate correlations between HF mesons (here we concentrate on D mesons) and light mesons, which offers a more accessible answer to the central question: "How much anisotropic flow does HQ gain from the bulk flow ?".

\section{Method}
We rely on the recently developed EPOSHQ model which combines the latest version of EPOS (EPOS3 \cite{EPOS3:2014}, now including a viscous fluid dynamical evolution of the bulk) with the MC@sHQ model described in~\cite{Nahrgang:2015}. In this study, dedicated to flow observables, only the case of pure elastic HQ energy loss (with K=1.5 cranking factor) will be considered. Preliminary results of MC@sHQ with 
EPOS3 background indicate a fair comparison wrt STAR data. The main developments implemented in EPOSHQ are a) HQ production described consistently in the EPOS initial conditions based on the semi-hard pomeron framework and b) HF mesons reinteraction in the hadronic phase using the URQMD afterburner. For the purpose of clear cut analysis, D mesons will be assumed to be produced solely from c quarks fragmentation. 
The analysis will be performed EBE using the event plane method (other methods will be investigated in an upcoming publication). We have also decided to crank up artificially the production of c quarks by a factor 50 in order to reduce the $v_n$ fluctuations resulting from the limited number of c quarks~\cite{Poskanzer:1998}. In practice this is achieved by cumulating 50 so-called "HQ events" for a single EPOS3 global event. Experimentally, this oversampling could be achieved by accumulating D mesons stemming from 50 events with similar $v_n$ of light particles. 
We only present results for $\sqrt{s}=2.76~{\rm TeV}$ PbPb collisions, due to limited space, but expect similar conclusions for $\sqrt{s}=200~{\rm GeV}$ and 5.02~TeV as well.

\section{Results}
As it is by now clear that the event plane method suffers from possibly large non-flow contributions if the event plane is extracted from the same particle ensemble as the one used to extract the magnitude of the flow harmonics $v_n$, we group, for each event, particles in 2 bins according to $|y|>2$ and $|y|<2$ for the $\hat{\Psi}_n^{\rm EP}$
and $\hat{v}_n\{\rm EP\}$ evaluation respectively\footnote{according to $\hat{\Psi}_n^{EP}=\frac{1}{n}{\rm atan}^{-1}\frac{\sum_{i=1}^{N} \sin n\phi_i}{\sum_{i=1}^{N} \cos n\phi_i}$ and $\hat{v}_n\{\rm EP\}=\frac{1}{N}\sum_{i=1}^N \cos(n(\phi_i-\hat{\psi}_n^{\rm EP}))$, where $N$ is the number of selected particles}, where the "hat" symbol refers to quantities evaluated in a single event. 
\begin{figure}[H]
\centering
\includegraphics[width=0.8\textwidth]{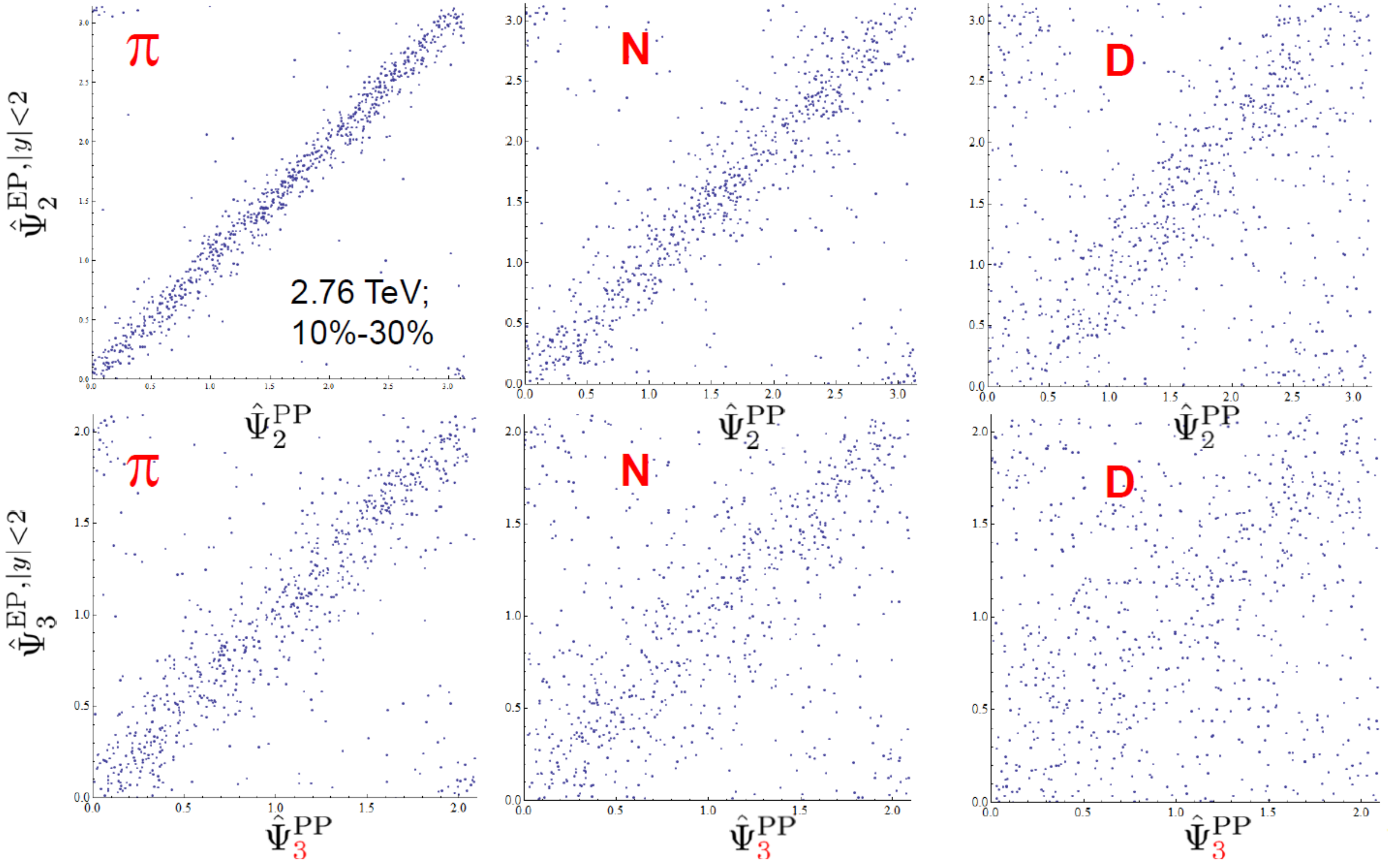} 
\caption{Correlation between the participant plane and the event plane for various types of particles in the 10\%-30\% centrality class of $\sqrt{s}=2.76{\rm TeV}$ PbPb reactions.}
\label{fig:1}
\end{figure}
We have extracted the event-planes $\hat{\Psi}_n^{\rm EP}$ from $|y|>2$ pions and found a nice correlation -- not shown here -- with the participant plane (PP) evaluated from the initial pomeron distribution preceeding QGP phase and thus identical for all types of light particles ensuing from QGP freeze out. We can thus use each of these quantities as a reference for the main direction of bulk expansion, and PP will be favored for this purpose.
In fig. \ref{fig:1}, we hence show the correlation between the PP and the event plane extracted for various types of particles produced with $|y|<2$. We observe a very good correlation for pions (both in the $n=2$ and $n=3$ harmonics). This correlation, however, suffers from larger and larger fluctuations with increasing mass of the particles (especially in the $n=3$ harmonics). One should nevertheless notice that this increase of the fluctuations is {\it not due} to a mere reduction of the number of particles in the sample. Indeed, our oversampling of c-quarks leads to twice as much D mesons as protons in the final state. We therefore conclude that these large fluctuations seen between  $\hat{\Psi}_n^{\rm PP}$ an $\hat{\Psi}_n^{\rm EP}(D)$ are a possible sign of an uncomplete thermalization of c quarks in the QGP.

\begin{figure}
\centering
\includegraphics[width=0.8\textwidth]{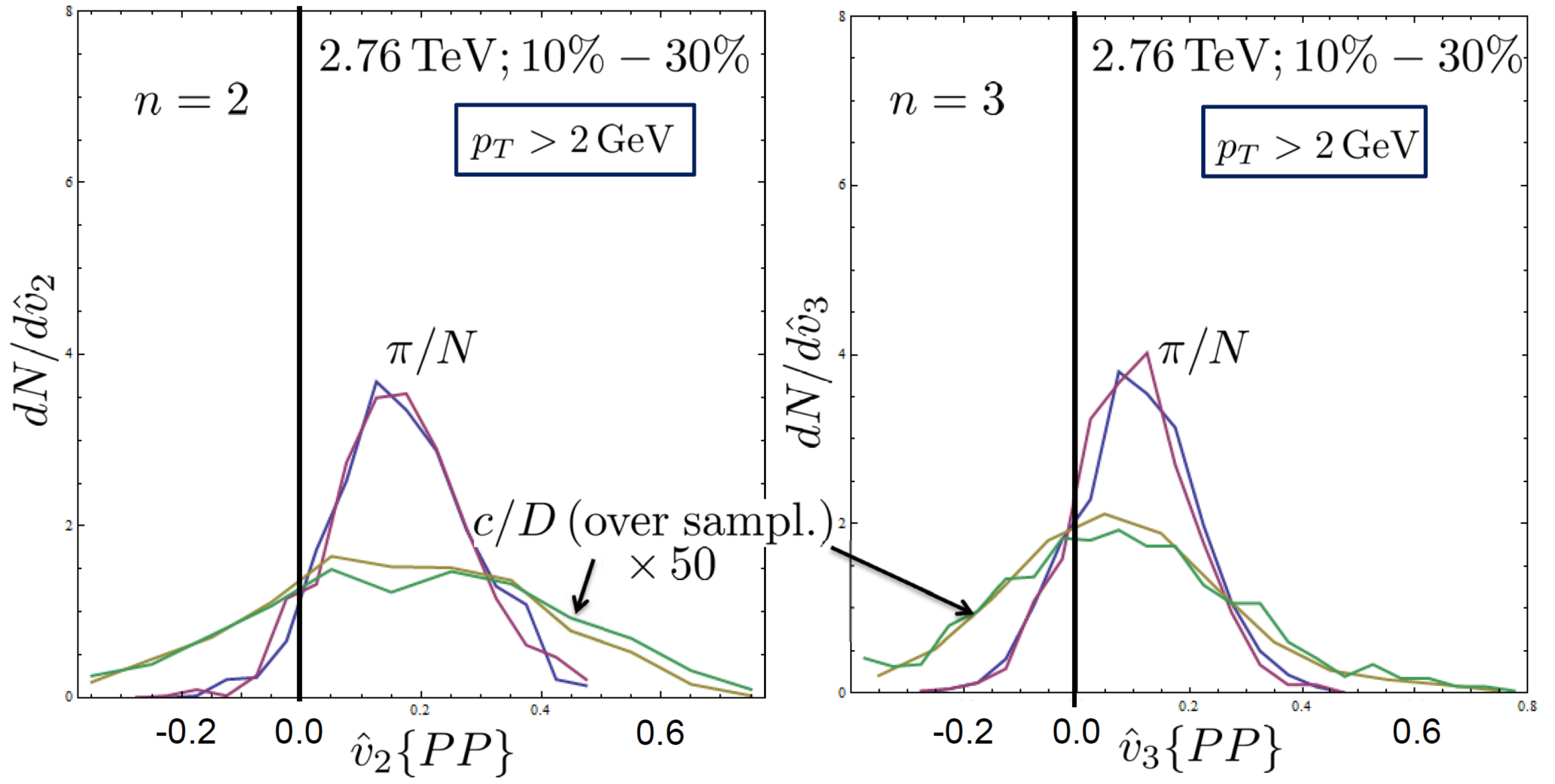} 
\caption{Left: Distribution of $\hat{v}_2$ for $\pi$, p, c quarks,
(before hadronization) and D mesons; Right: same for $\hat{v}_3$.}
\label{fig:2}
\end{figure}

In fig.~\ref{fig:2}, we show the distributions of the $\hat{v}_{n=2,3}$ obtained for $\pi$, p, c-quarks (before hadronization) and 
D mesons obtained taking $\Psi_n^{\rm PP}$ as the reference plane for all species. We apply an extra $p_T>2\,{\rm GeV}$ cut off in order to focus on intermediate $p_T$ D mesons observed experimentally.
We see that all species roughly share the same average for $n=2$, while $\langle \hat{v}_3(D) \rangle\approx\langle\hat{v}_3(c)  \rangle< \langle\hat{v}_3(\pi) \rangle\approx \langle\hat{v}_3(p) \rangle$, indicating on the average a finite but limited coupling of c quarks to the bulk flow. However, $\hat{v}_n$ of c quarks and D mesons suffers from strong fluctuations, characterized by standard deviations larger than averages. D mesons are in fact found rather often with a $\hat{v}_n$ value with {\em opposite sign} to the bulk one. We insist that these large fluctuations cannot be attributed primarily to the small number $N_c$ of c quarks, as we have $N_c\approx 2 N_\pi \approx 2 N_p$ for $p_T>2\,{\rm GeV}$, due to the $\times50$ oversampling. We also observe this phenomena if no
$p_T$ cut is applied (except that the standard deviation for the $\hat{v}_n(\pi)$ gets strongly reduced), independent of the centrality class which is investigated. In passing we note that the hadronic rescattering in URQMD has little influence on these observables.

\begin{figure}[H]
\centering
\includegraphics[width=0.8\textwidth]{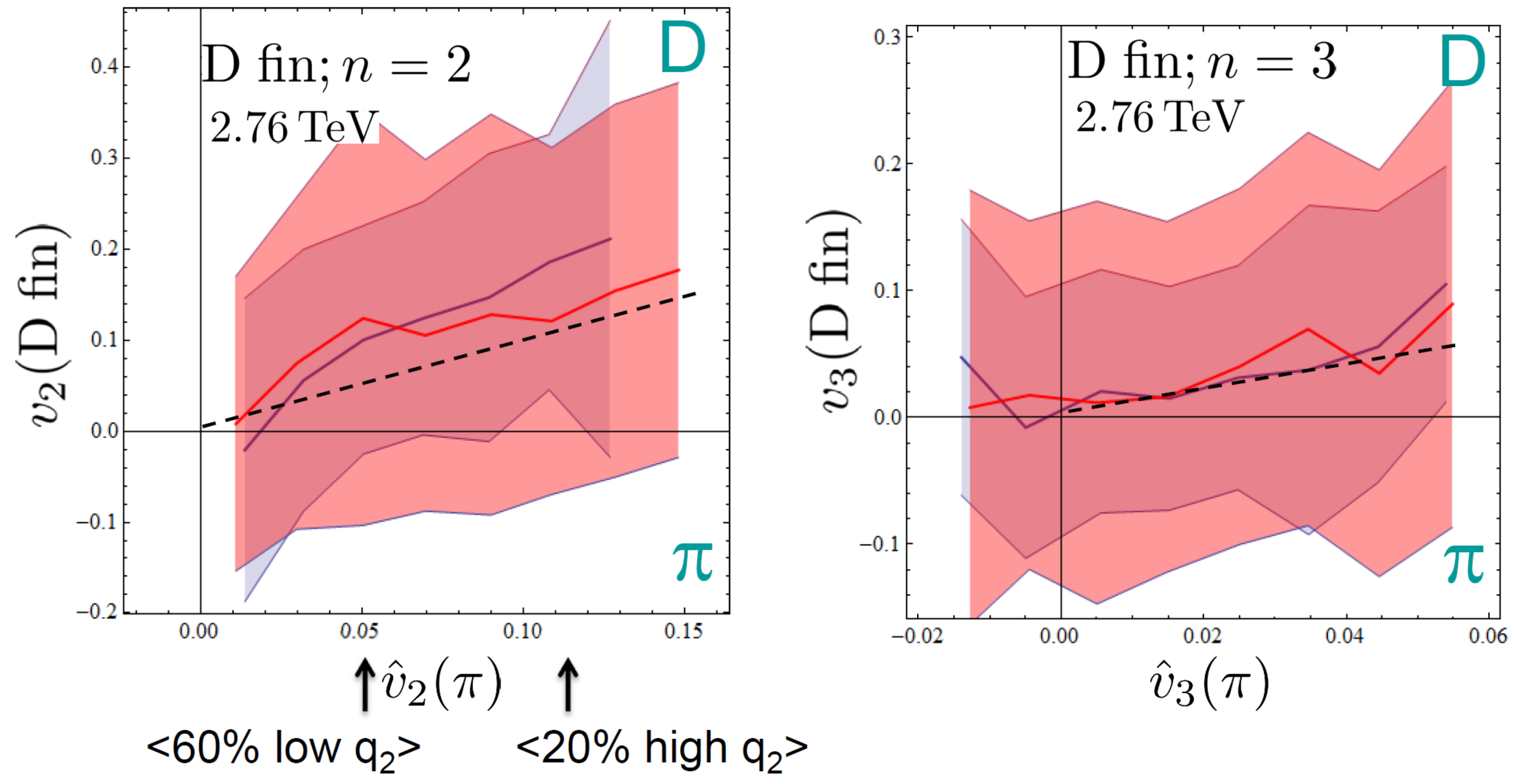} 
\caption{Left (color online): $v_2$ of D mesons (average value, average value $\pm$ 
standard deviation) as a function of $\langle \hat{v}_2(\pi) \rangle$ for 10\%-30\% (blue) and 30\%-50\% (red) centrality classes; the dashed line represents case of equal average elliptic flow; Right: same for $v_3$.}
\label{fig:3}
\end{figure}

In fig. \ref{fig:3}, we investigate the correlations between 
$\hat{v}_n$ of D mesons and pions, for 10\%-30\% and 30\%-50\% simultaneously. For this purpose, we bin the 
$\hat{v}_n(\pi)$ and then evaluate the average and standard deviations of $\hat{v}_n(D)$ in each of these bins, as could be done experimentally. Due to our present limited statistics, we did not apply any cut off on $p_T$. We see that, irrespective of the centrality class, the D mesons $\langle \hat{v}_2\rangle$ correlates linearly with $\langle \hat{v}_2\rangle$ of pions at small values ($\lesssim 0.05$) but shows  some saturation for larger values. For $v_3$ a linear rise is found up to the largest values. Although precise trends might slightly differ if a $p_T$ cut off is applied, one can conclude that the (average values of the) flow harmonics are well correlated. 
We also conclude from this analysis that the fluctuations of $\hat{v}_n$ are slightly larger for the 30\%-50\% centrality class, an effect that could be simply explained by the less numerous production of c quarks in this class: $N_c(30\%-50\%)\approx 0.4 N_c(10\%-30\%)$. Fluctuations are also found to be rather independent of the $\langle \hat{v}_2(\pi)\rangle$ bin which points to the fact that the EBE eccentricities $\epsilon_n$  play a large role for the average flow harmonics but not for the standard deviations. We want also mention that  in \cite{Prado:2016szr} -- where a similar analysis was performed --  smaller fluctuations for the same observable have been observed, what might be attributed to much larger oversampling than our "$\times 50$" prescription. One should however refrain from drawing premature conclusions as the $p_T$ range of selected D mesons differ between~\cite{Prado:2016szr} and this study.

\section{Discussion and Conclusion}
The conclusion from our study is somehow ambiguous. On the one side, if we consider average values only, one would conclude that c quarks with small $p_T$ ($p_T\approx m_c$) benefit strongly from bulk flow and thus "thermalize" with the medium. On the other side, one observes fluctuations of $\hat{v}_{2,3}(c)$
and $\hat{v}_{2,3}(D)$ that exceed standard expectations \cite{Poskanzer:1998} (even when EBE fluctuations of the initial $\epsilon_n$ are taken into account), which should be considered as a signal for lack of complete thermalization. In this respect, one should recall that light hadrons and HF hadrons have quite different origins in our modeling: While light hadrons with small $p_T$ are exclusively produced on the freeze out hypersurface according to statistical hadronization~\cite{EPOS:2010}, HF mesons are the final outcome of c quarks created at initial stage with a 
non thermal $p_T$ distribution and possibly large specific EBE fluctuations. Further investigations will be mandatory in order to understand to which extend the $\hat{v}_n$ distribution of c quarks prior to their hadronization into D mesons is impacted by the evolution of these quarks in the QGP phase, as well as the precise impact of the oversampling adopted in this study. In all cases, we conclude that correlations of heavy-light flow harmonics have very promising perspectives and encourage experimental collaborations to further investigate these observables.

\section*{Acknowledgement}
We gratefully acknowledge the support from the TOGETHER project, R\'egion Pays de la Loire (France) as well as enlightening discussions
with J.Y. Ollitrault.



%% References


% \bibliographystyle{elsarticle-num}
% \bibliography{<your-bib-database>}



\end{document}